\documentclass[12pt]{iopart}
\usepackage{graphicx}
\usepackage{xcolor}
\usepackage{caption}
\usepackage{subcaption}
\usepackage[normalem]{ulem}
\begin{document}
\title[Time-resolved neutron isotropy in a SFS z-pinch]{Time-resolved measurement of neutron energy isotropy in a sheared-flow-stabilized Z pinch}

\author{R. A. Ryan$^1$, P. E. Tsai$^1$, A. R. Johansen$^3$, A. Youmans$^2$, D. P. Higginson$^2$, J. M. Mitrani$^2$, C. S. Adams$^1$, D. A. Sutherland$^1$, B. Levitt$^1$, U. Shumlak$^{1,3}$}

\address{$^1$ Zap Energy, Everett, WA 98203 USA}
\address{$^2$ Lawrence Livermore National Laboratory, 7000 East Avenue, Livermore, CA 94550 USA}
\address{$^3$ Aerospace and Energetics Research Program, University of Washington, Seattle, WA 98195 USA}

\begin{abstract}
Previous measurements of neutron energy using fast plastic scintillators while operating the Fusion Z Pinch Experiment (FuZE) constrained the energy of any yield-producing deuteron beams to less than 4.65~keV. FuZE has since been operated at increasingly higher input power, resulting in increased plasma current and larger fusion neutron yields. A detailed experimental study of the neutron energy isotropy in these regimes applies more stringent limits to possible contributions from beam-target fusion. The FuZE device operated at $-25$~kV charge voltage has resulted in average plasma currents of 370~kA and D-D fusion neutron yields of $4\times10^7$ neutrons per discharge. Measurements of the neutron energy isotropy under these operating conditions demonstrates the energy of deuteron beams is less than $7.4 \pm 5.6^\mathrm{(stat)} \pm 3.7^\mathrm{(syst)}$~keV. Characterization of the detector response has reduced the number of free parameters in the fit of the neutron energy distribution, improving the confidence in the forward-fit method. Gamma backgrounds have been measured and the impact of these contributions on the isotropy results have been studied. Additionally, a time dependent measurement of the isotropy has been resolved for the first time, indicating increases to possible deuteron beam energies at late times. This suggests the possible growth of $m$=0 instabilities at the end of the main radiation event but confirms that the majority of the neutron production exhibits isotropy consistent with thermonuclear origin. 

\end{abstract}

%
%
%
%

\submitto{\NF}
\maketitle

\section{Introduction}

Fusion experiments utilizing the ``pinch effect'' date back to 1958 when neutron yields as high as $10^8$ neutrons per discharge were produced \textemdash a seemingly promising step towards the goal of producing controlled thermonuclear energy \cite{Anderson}. However, investigations of the neutron yield from these experiments revealed the dominating physics to be of beam-target origin, arising from an $m$=0 instability in the plasma \cite{Anderson}. This ``sausage'' instability features an axisymmetric perturbation of the azimuthal magnetic field, causing the growth of a local axial electric field that accelerates ions across the ``sausage links". In these early experiments, deuteron beam energies of up to 200~keV were measured. Additionally, $m$=1 ``kink modes'' led to a rapid deterioration of the pinch, significantly reducing plasma confinement times \cite{Kadomtsev}. 

\begin{figure}
    \centering
    \includegraphics[width=1.00\linewidth]{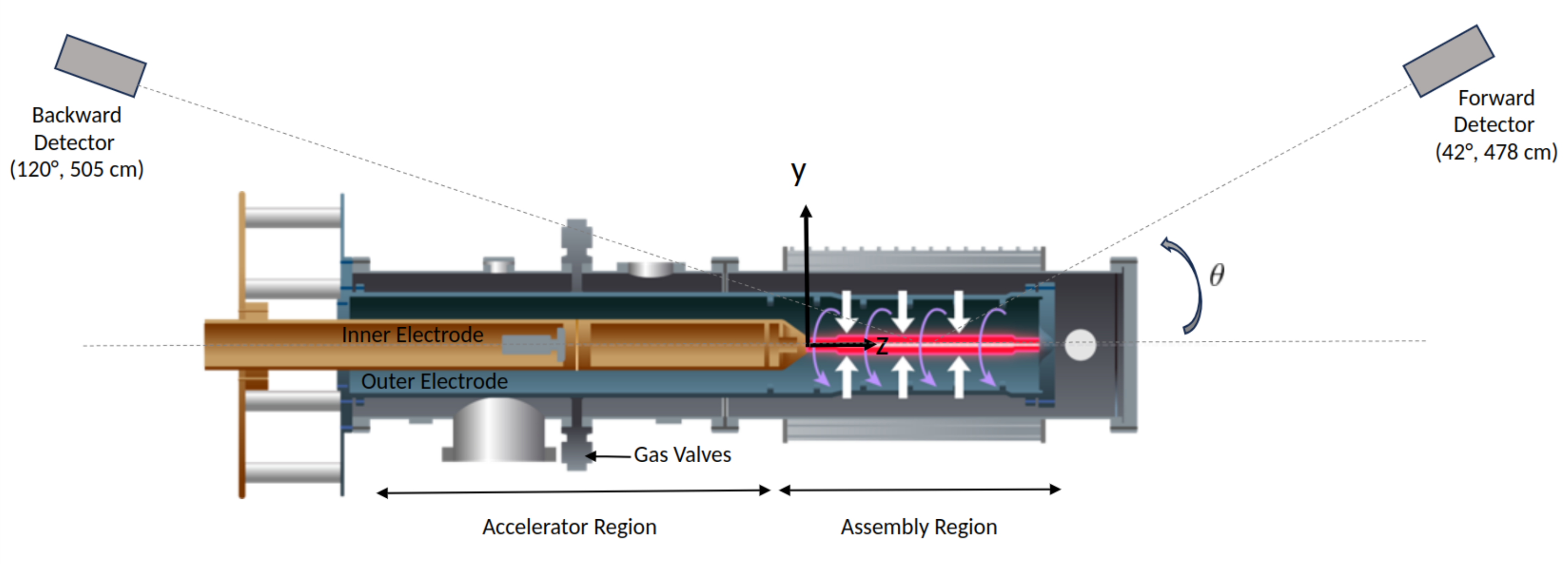}
    \caption{Diagram of the FuZE device. Gas is injected into the accelerator region where it is ionized. Ramp up of the current causes the plasma to accelerate and flow into the formation section and form a flowing Z pinch. The coordinate system is defined as shown, with the origin at the tip of the nosecone. The two scintillator detectors are shown with the angle to the z axis and radial distance from the nosecone.}
    \label{fig:fuze}
\end{figure}

Decades later, numerical simulations demonstrated that $m$=1 instabilities could be stabilized with a radially sheared axial flow \cite{Shumlak_Hartman} and initial experiments with shear flow stabilized (SFS) Z pinches demonstrated apparent stability maintained for hundreds of times longer than characteristic Ideal-Magnetohydrodynamics (MHD) instability growth times \cite{Shumlak:2001,ShumlakJAP2020,Zhang2019}. The FuZE device at Zap Energy produces SFS Z pinches at plasma temperatures relevant to thermonuclear fusion \cite{Levitt2024PRL}. As depicted in Figure \ref{fig:fuze}, gas is injected into a 20~cm diameter, 1-meter-length coaxial accelerator. A 100~kJ-class capacitor bank begins the discharge by ionizing the gas, then accelerates the resulting plasma within a coaxial electrode assembly, and produces a pinch in the assembly region. 

The mid-20th century history of pinch experiments showcased the importance of measuring potential beam-target contributions to neutron production, and the first neutron energy isotropy measurements made on the FuZE device demonstrated deuteron beam energies of less than $4.65$~keV \cite{Mitrani:2021}. These measurements were recorded while operating FuZE with a $-6$~kV bias voltage from a 180~kJ bank, producing pinch currents of 200~kA and neutron yields of $10^5$ per discharge with 80\% helium and 20\% deuterium by gas valve puff pressure. Recently the FuZE device has been operated with a slightly lower energy bank but four times higher voltage, resulting in more than twice the plasma current and neutron yields in the mid $10^7$ per discharge with 100\% deuterium gas injection, motivating a new measurement of the neutron energy isotropy. 


This paper describes the development of the isotropy analysis and results from higher performance regimes in the FuZE device, which bound the beam energy as a function of time during the discharge. The organization of this paper is as follows. Section~\ref{Methods} reviews the principles of measuring neutron energy anisotropy to deduce deuteron beam energy and describes new development of the analysis approach. Section~\ref{Results} highlights the time-resolved isotropy results for the $-25$~kV charge voltage and details the quantification of the two most prevalent sources of systematic error in the final result. Section~\ref{Conclusion} contains concluding remarks.

\section{Methods} \label{Methods}

\begin{figure}
    \centering
    \includegraphics[width=\linewidth]{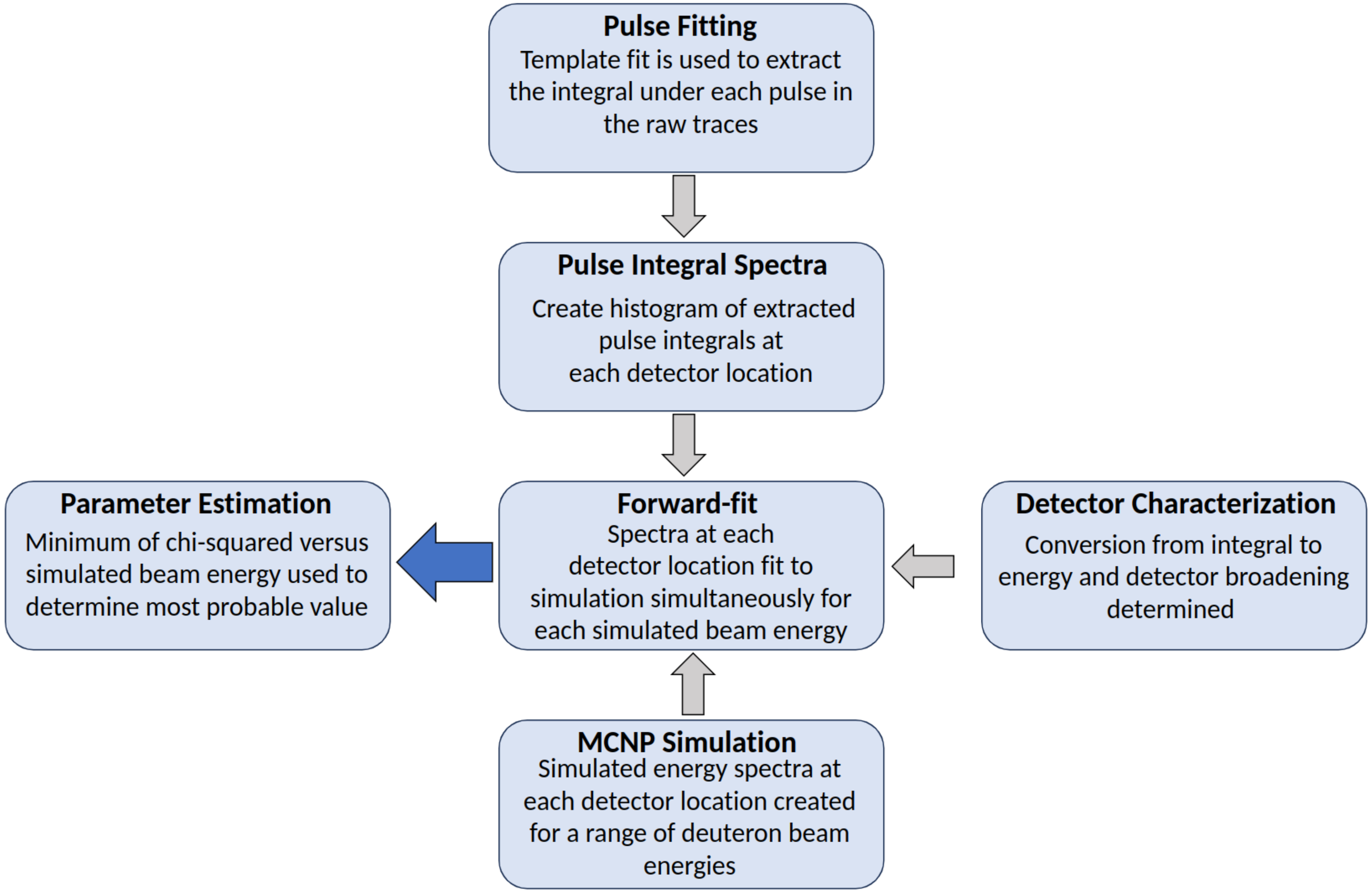}
    \caption{Summary of the methodology used to determine the most probable deuteron beam energy from measured neutron energy anisotropy. Pulse integral spectra created from fitting individual pulses at each detector location are fit to simulated spectra produced in MCNP. Several fit parameters are fixed based upon the detector characterization, and a chi-squared minimization of the forward-fit is used to extract the value and statistical error in the observed deuteron beam energy. }
    \label{fig:analysis-flowchart}
\end{figure}

To measure the energy isotropy of the neutron emission, plastic scintillator detectors are deployed at two locations relative to the center of the FuZE device and the shift in the observed energy distributions between the two detector systems is used to determine the most probable deuteron beam energy by fitting to spectra simulated via the Monte Carlo N-Particle code (MCNP) \cite{TechReport_MCNP}. The general steps to this methodology are illustrated in the flowchart of Figure \ref{fig:analysis-flowchart}. 

\subsection{Experimental setup}

For the FuZE device, ion beams would propagate in the $-z$-direction (coordinate system shown in Figure~\ref{fig:fuze}). Ideally, the sensitivity to the energy anisotropy is maximized by placing detectors at $0^{\circ}$ and $180^{\circ}$ relative to this assumed beam axis. However, minimization of scattering is desired in order to reduce error, constraining the detector array placements to lines of sight at angles of $42^{\circ}$ and $120^{\circ}$ relative to the pinch axis, as illustrated in Figure~\ref{fig:fuze}. Each detector array is comprised of scintillators which measure the neutron energy distribution. 

\subsubsection{Plastic scintillator detectors}
Implementations of plastic scintillating detectors on the FuZE device for measuring neutron isotropy have been discussed previously \cite{Mitrani:2021}. To adjust for the increased neutron yield with recent FuZE operating conditions, the scintillator volume has been reduced compared to the previous measurement. Each detector assembly consists of a 10~mm~$\times$~10~mm right circular cylinder of EJ-230 plastic scintillator optically coupled to an H1949-51 Hamamatsu photomultiplier tube (PMT) and digitized by a 1~GHz bandwidth, $3.125$~GS/s MSO58LP Tektronix oscilloscope, which resolves individual pulses produced by neutrons elastically scattering within the scintillator material.

\subsubsection{Organic glass scintillators}

The plastic scintillators used for the neutron isotropy measurement are also sensitive to gamma radiation, but signals produced by neutrons and gammas are experimentally indistinguishable in the EJ-230 plastic. The simulated spectra modeled by MCNP include gamma interactions. To verify that the simulated gamma background is adequately modeled, an additional set of organic glass scintillator detectors from BlueShift Optics, capable of distinguishing between recoil and electron excitation events, has been fielded in locations next to the plastic scintillators \cite{CARLSON2016}.

\subsection{Detector response characterization} \label{Characterization}

Two important values are required to characterize the detector response: the calibration from units of charge to electron equivalent energy and the Gaussian energy broadening of the energy spectra due to the finite detector resolution. The method of quantifying the light output and resolution of scintillator materials using gamma sources of known energy has been well established \cite{SICILIANO2008} \cite{DIETZE1981}. 

An MCNP simulation consisting of the basic detector geometry located 5 cm from a gamma source was run for Cs-137, Mn-54, and Co-60 sources. Experimental pulse energy distributions, produced by exposing the scintillators to each gamma source, were fit to the simulated distributions to simultaneously extract the calibration factor and Gaussian energy broadening.
Calibration factors mapping scintillator light output from gammas, in units of MeVee, to proton recoil energy, in units of MeV, have been obtained by several methods \cite{pozzi,weldon,manfredi}. While all three literature calibrations produced consistent results, the results of Weldon et al. resulted in the best forward-fits and are used for the analysis presented here \cite{weldon}.

\subsection{MCNP simulation} \label{MCNPSim}

The full FuZE geometry including materials within the vacuum vessel, large neutron scattering objects in the room, and key components of the detector geometry at the two prescribed locations were simulated with MCNP6 to model the energy deposition of each neutron inside the scintillators. The motivation for modeling the inner geometry of the FuZE device and surrounding room materials was twofold: to simulate the effects of neutron scattering and to estimate prompt gamma radiation produced via neutron interactions with the surrounding material. Both background sources could impact the resolution of the measured neutron energies. Gamma contributions to the scintillator signal are saved to the output files along with the neutron interactions and are included in the analysis described in subsequent sections.

A series of source energies, consisting of a line source of neutrons with energies and emission angles characteristic of a specific deuteron beam energy, $E_d$, were simulated. The magnitude and direction of $E_d$ was varied from -50~keV to 50~keV in 2.5~keV increments. The energy deposition of recoiled protons and Compton electrons in the scintillators was saved for each simulation. 



Calibration factors mapping scintillator light output from gammas, in units of MeVee, to proton recoil energy, in units of MeV, have been obtained by several methods \cite{pozzi,weldon,manfredi}. While all three literature calibrations produced consistent results, the results of Weldon et al. resulted in the best forward-fits and are used for the analysis presented here \cite{weldon}.

\subsection{Pulse fitting and pulse energy spectra creation}

\begin{figure}
\centering
\includegraphics[width=\textwidth]{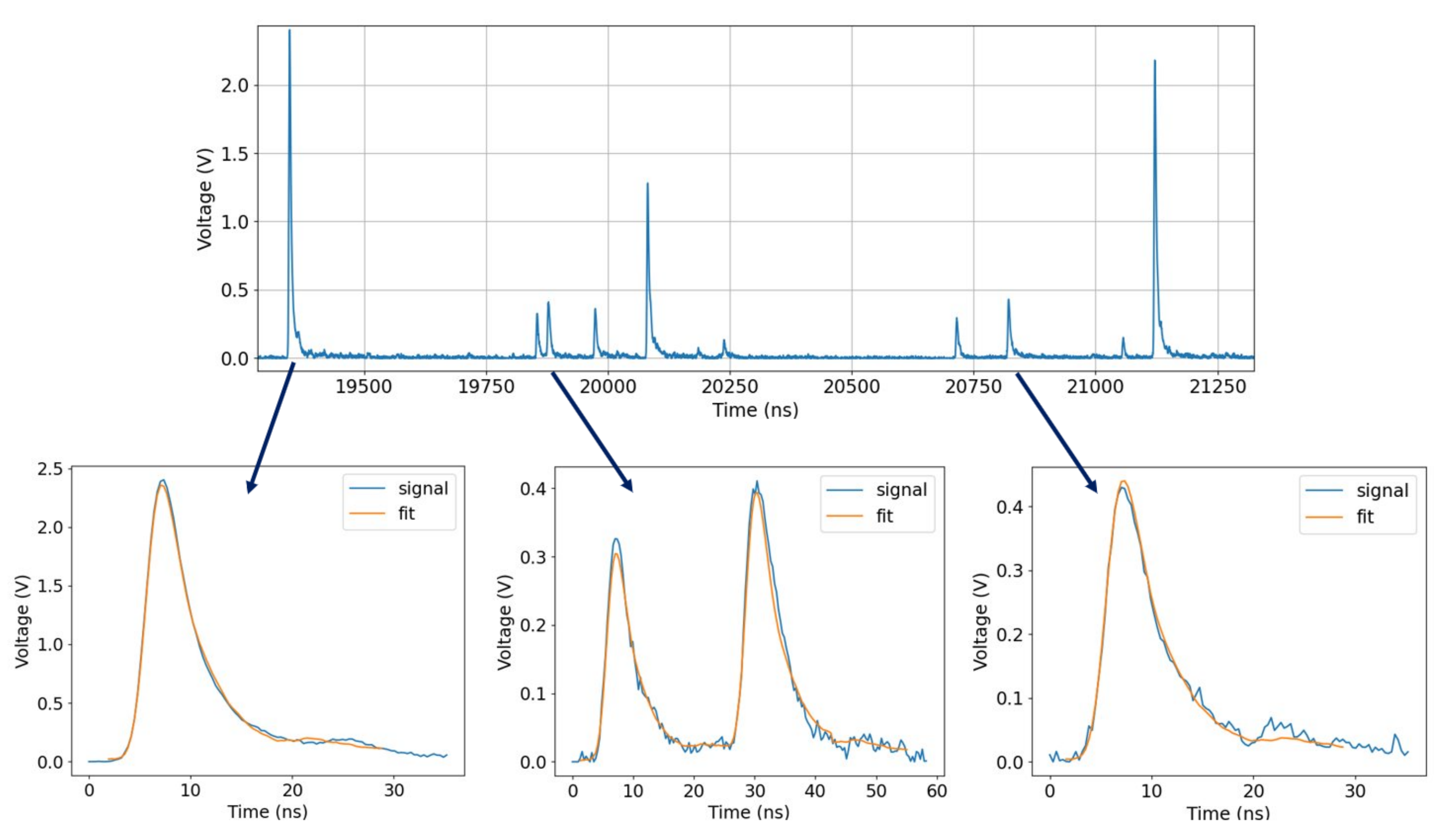}
\caption{A raw digitized trace from a single detector for one FuZE discharge (top) and an example of the fits to several pulses within the trace. For pulses which are overlapping within the 28~ns transit time of the PMT, the template allows for a simultaneous fit to extract the energy of both pulses.}
\label{pulseFit}
\end{figure}

An example of the raw PMT signal recorded during a single Z-pinch plasma discharge is shown in Figure~\ref{pulseFit}. In previous analyses, an analytical function was used to fit each pulse \cite{Mitrani:2021}. To improve the fit for overlapping pulses, the fitting has been changed from an analytical function to a template fit which uses the average pulse shape and scales in width and height to fit each pulse.

The integral of each pulse is obtained by integrating the area beneath the fitted template and converting to units of MeV using the calibration method outlined in Section~\ref{Characterization}. The determined energy of each pulses is added to a histogram, resulting in an energy spectrum of all pulses at each detector location.

\subsection{Forward-fit}\label{fit_analysis}

The spectra of pulse energies recorded at the two detector locations are fit to MCNP-simulated spectra to determine the most likely beam energy. Previously, the analysis fit spectra recorded at each detector location separately and contained six free-floating parameters: a scalar to convert the measured charge to electron equivalent energy (MeVee), three parameters to characterize the energy dependent detector resolution, and a normalization scalar for each detector \cite{Mitrani:2021}. In order to better constrain the forward-fit method, the calibration and resolution parameters have been fixed using the values obtained from the detector characterization discussed in Section \ref{Characterization}. Additionally, the analysis has been developed to fit both detectors simultaneously.

The simulated number of counts in the $i$th bin of the detector at location $j$ for a given deuteron beam energy is given by

\begin{equation}
\tilde{N}_\mathrm{sim}^j(i;E_d) = A_j N_\mathrm{sim}^j(i;E_d) \ast R_j(\sigma)
\end{equation}

\noindent where $A_j$ is a free fit parameter that adjusts the scale of the simulated data, and the spectrum is broadened via convolution with a detector-dependent resolution function, $R_j$, determined by analyzing the calibration spectrum. The number of observed counts in the $i$th bin at each detector location, $j$, is given by

\begin{equation}
 \tilde{N}_\mathrm{obs}^j(i) = N(C\cdot i)
\end{equation}

\noindent where $C$ is a calibration adjustment factor to allow for 10\% variation of the calibration from V$\cdot$ns to MeVee. For the upstream-downstream detector pair, the three floating fit parameters $A_1$, $A_2$, and $C$ are determined for each simulated value of $E_d$ by regression, minimizing the chi-square statistic

\begin{equation}
S(E_d) = \sum_{j=1,2} \sum_{i} \frac{(\tilde{N}_\mathrm{sim}^j(i;E_d)-\tilde{N}_\mathrm{obs}^j(i))^2}{\tilde{N}_\mathrm{obs}^j(i)}.
\end{equation}

\section{Results and discussion} \label{Results}

\subsection{Measured beam energy from the pulse fitting results}

\begin{figure}
     \centering
     \begin{subfigure}[b]{0.45\textwidth}
         \centering
         \includegraphics[width=\textwidth]{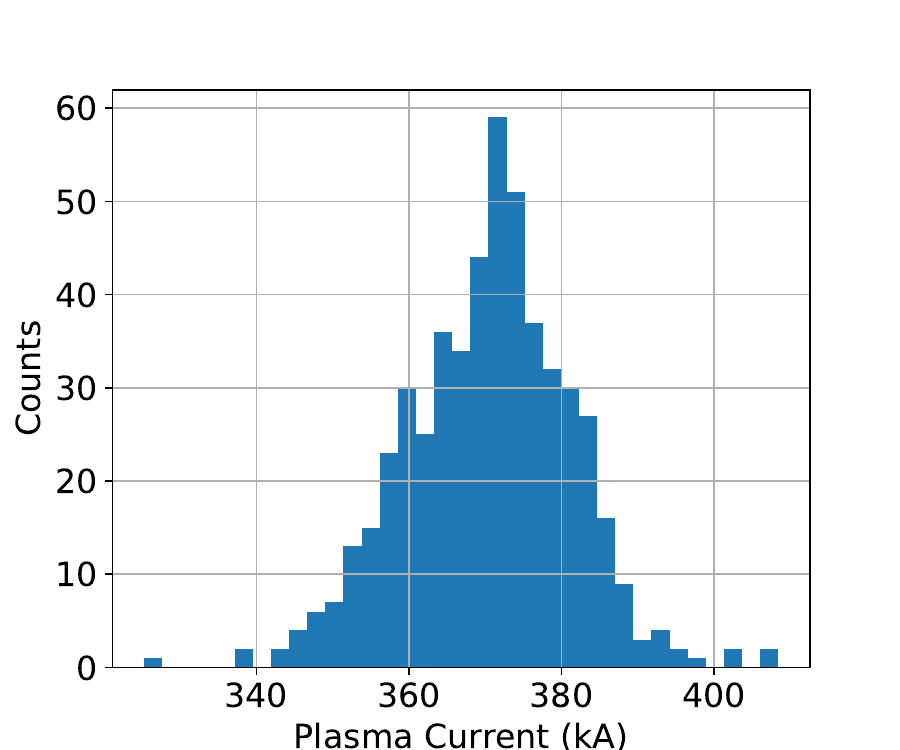}
         \caption{}
         \label{plasma_currents}
     \end{subfigure}
     \hfill
     \begin{subfigure}[b]{0.45\textwidth}
         \centering
         \includegraphics[width=\textwidth]{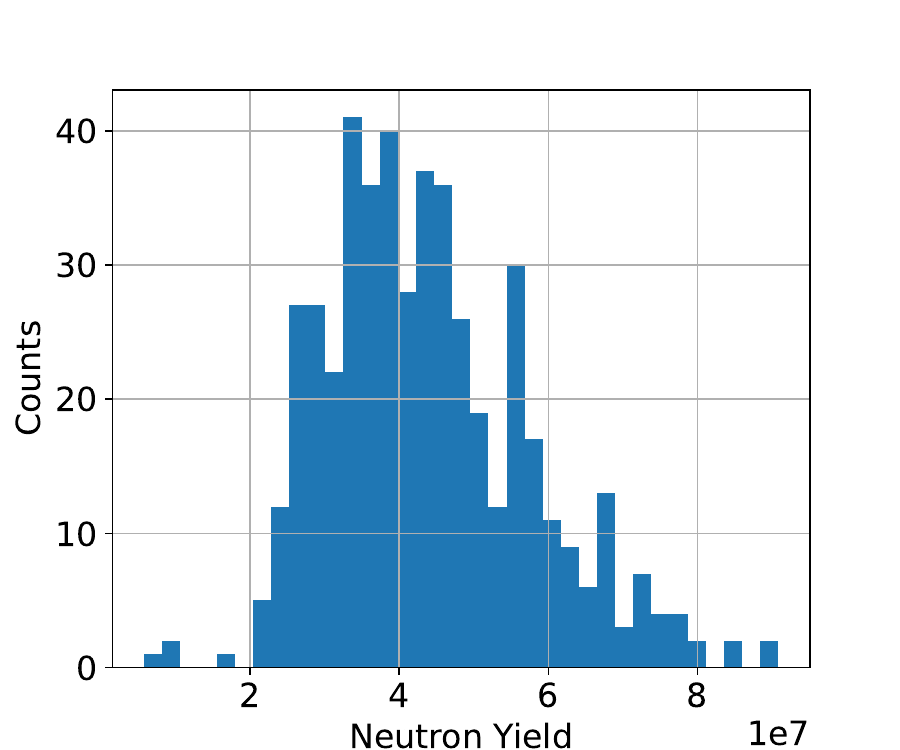}
         \caption{}
         \label{yields}
     \end{subfigure}
     \caption{(a) Distribution of (a) plasma current measured during radiation time and b) Neutron yield distribution demonstrating the reproducibility of the shot parameters.}
     \label{repeat_params}
\end{figure}

Previous measurements of neutron isotropy on the FuZE device determined any contributions from beam-target fusion were constrained to a deuteron beam energy below 4.56~keV for $-6$~kV operation \cite{Mitrani:2021}. For the present study, FuZE is supplied by a lower energy bank operated at higher charge voltage, resulting in higher plasma currents and three orders of magnitude greater neutron yield. The FuZE device was operated with fixed settings at a bank voltage of $-25$~kV for 433 discharges to generate a dataset with a high number of observed pulses. The distribution of the values of neutron yield and plasma current illustrating the repeatability of the discharge are displayed in Figure~\ref{repeat_params}.

Although there is some spread, both the plasma current and neutron yield from every discharge form a well-peaked distribution, with mean plasma current of 336~kA and neutron yield of $4\times10^7$. The neutron energy spectra from each detector location were summed across all discharges and the forward-fit method was applied to determine the extent of neutron isotropy. 

\begin{figure}

\centering
\includegraphics[width=.32\textwidth]{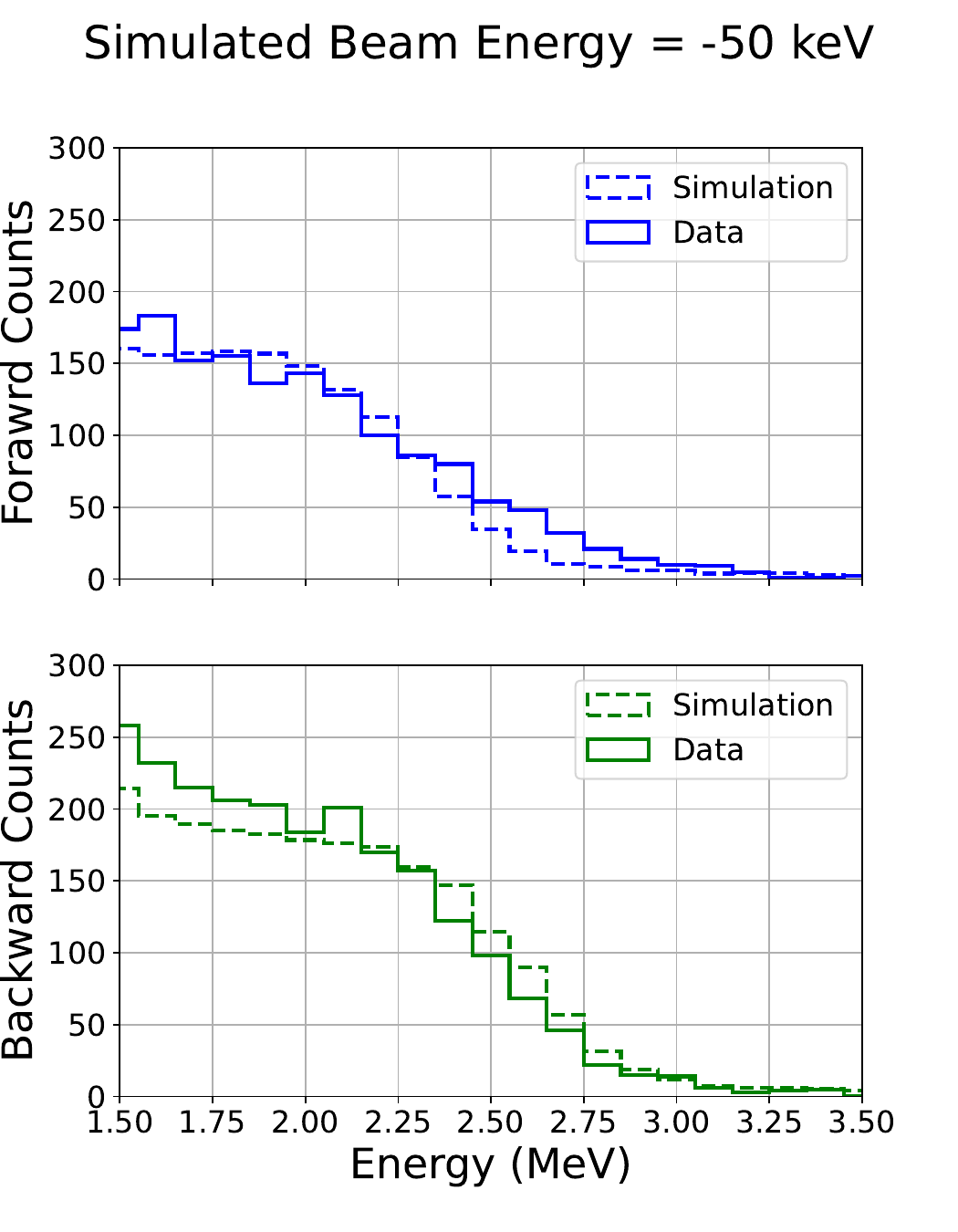}\hfill
\includegraphics[width=.32\textwidth]{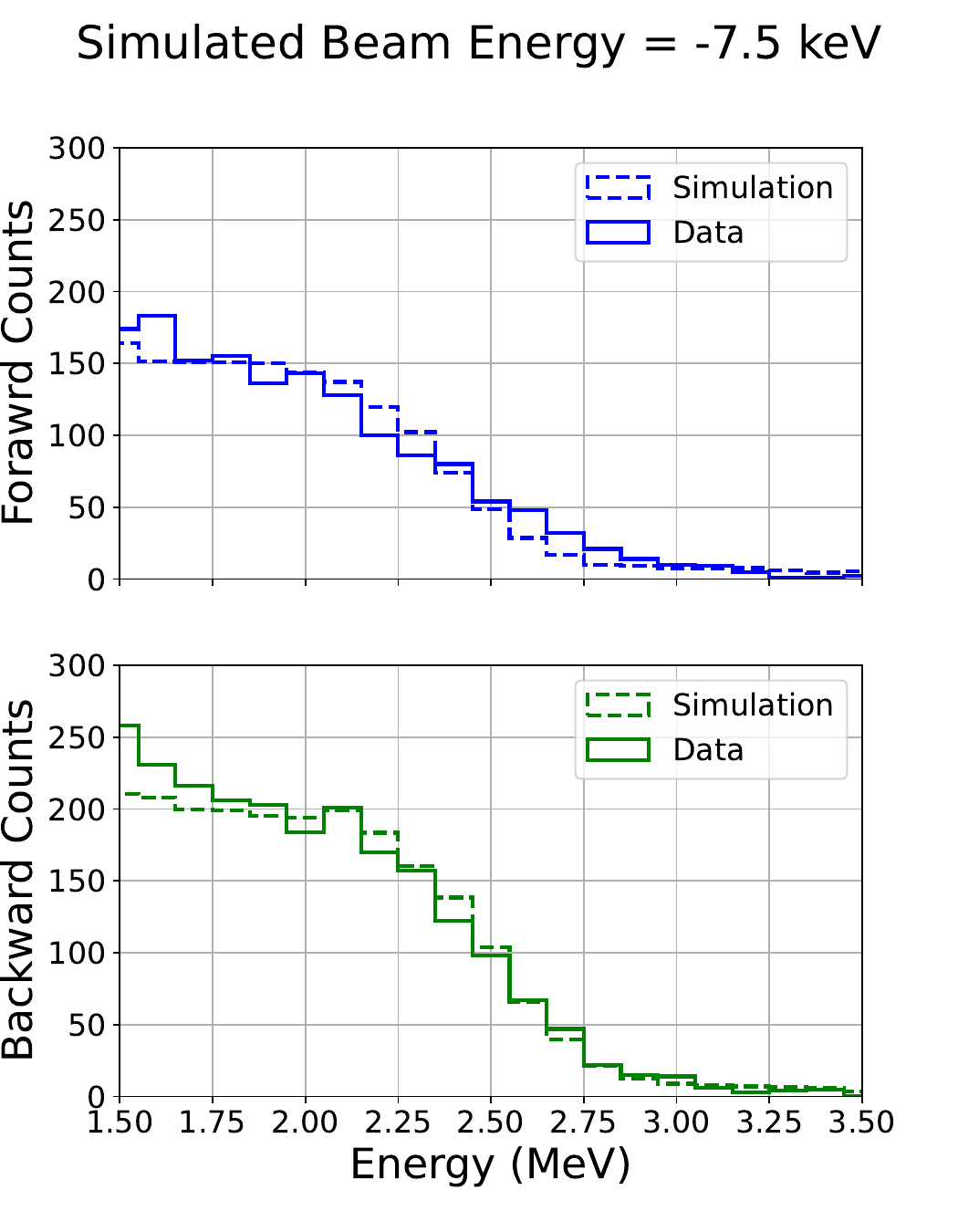}\hfill
\includegraphics[width=.32\textwidth]{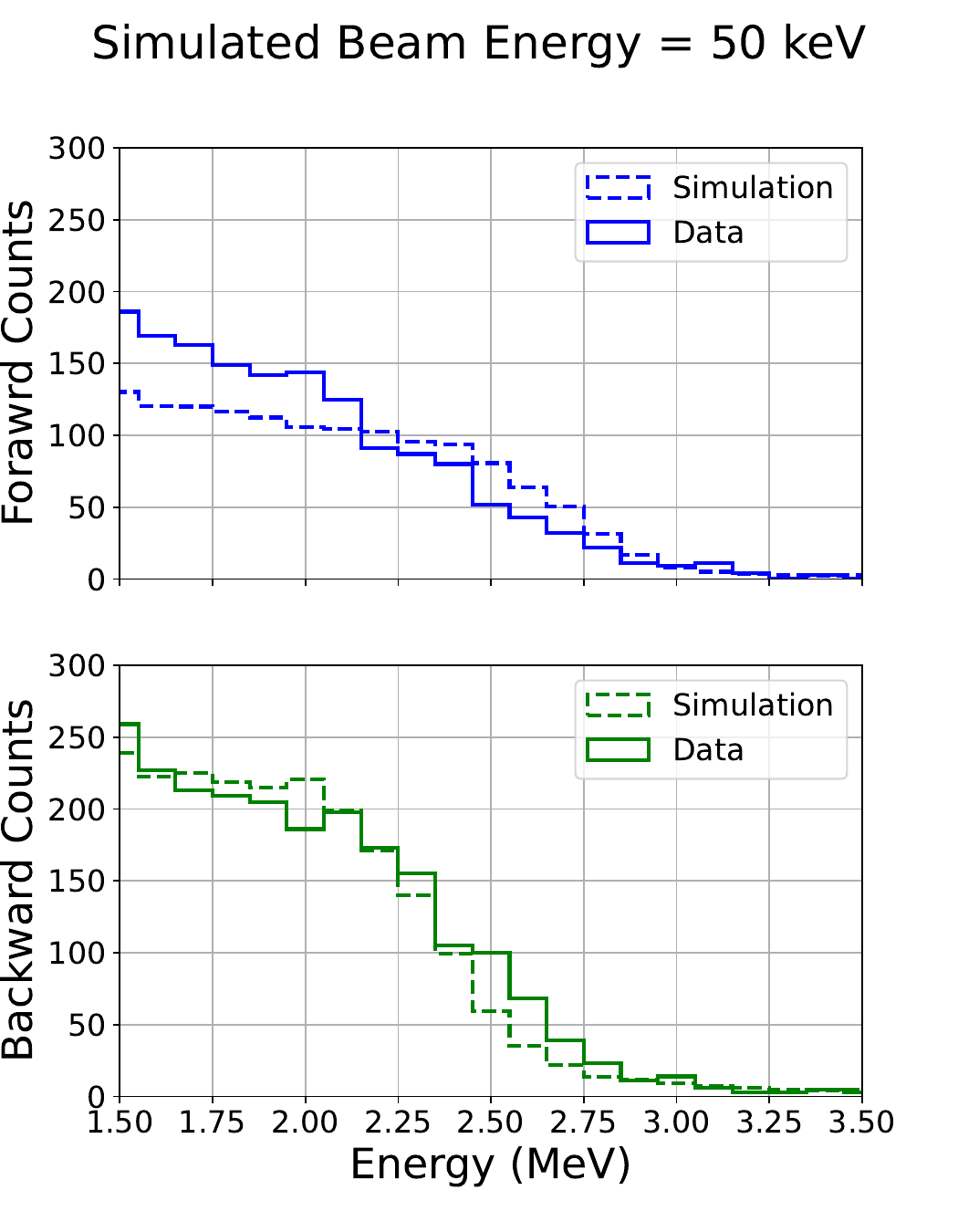}

\caption{Fits of the experimental data (solid) to the Gaussian energy broadened simulation data (dashed) for three values of the deuteron beam energy, $E_d$. The forward detector (top row) and backward detector (bottom row) are simultaneously fit to minimize the chi-square with three floating parameters. Deficiencies in the fit are easily observed for beam energies of 50~keV. Of the examples shown here, the data is best fit to the simulation with a beam energy of 7.5~keV in the -z direction. }
\label{mcnpfit}

\end{figure}

Best fits between the experimental spectra and simulation using the forward-fit are shown for three different simulated beam energies in Figure \ref{mcnpfit}. Each fit displays the experimentally measured spectra as a solid line and the simulated spectra as a dashed line, for both the forward and backward detectors. The chi-square statistic of each simultaneous fit to both the forward and backward distributions is determined from the three parameter fit discussed in section \ref{fit_analysis}. Shifted values of the edge location between the experimental and simulated spectra are observed for beam energies of $\pm 50$~keV, demonstrating the sensitivity of the fit. For comparison, the spectra with 7.5~keV ion beam energy in the negative $z$ direction displays the best agreement between the simulated and measured spectra for both detector locations.

\begin{figure}
\centering
\includegraphics[width=0.75\textwidth]{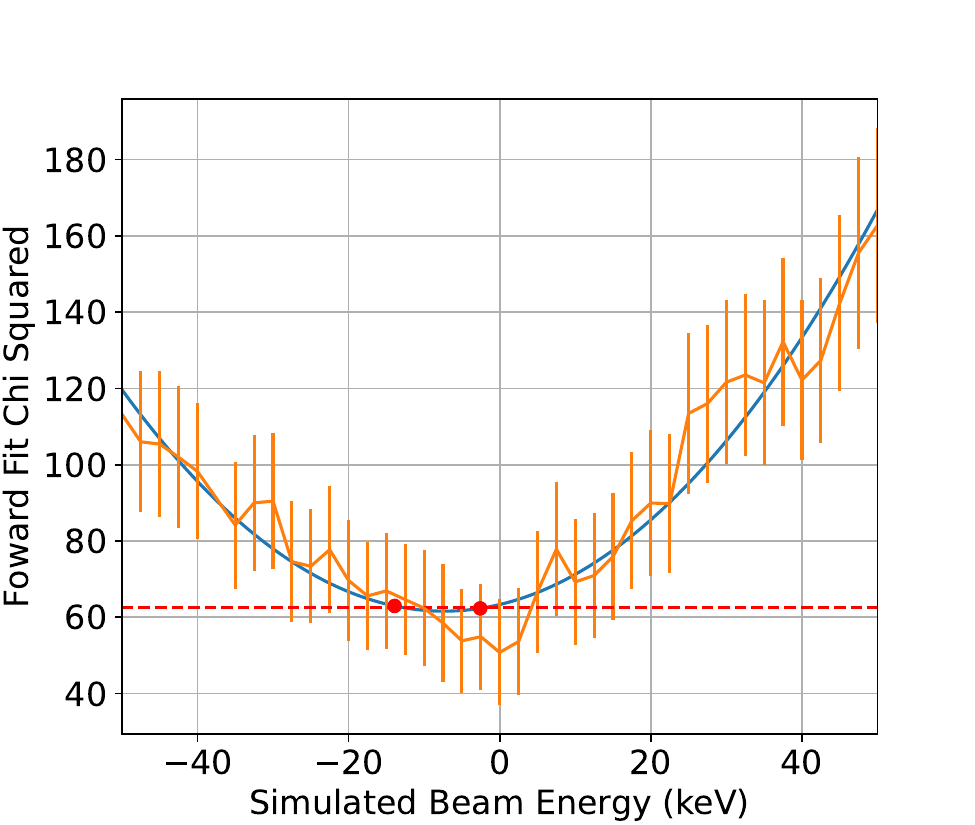}
\caption{Parabola fit to estimate the value of deuteron beam energy, $E_d$, which best matches the simulated spectra to the observed data. The best-fit value of $E_d$ is given by the value at the minimum and the uncertainty in this estimation is determined from the width of the parabola at the minimum chi-square statistic plus 1, denoted by the dashed line.}
\label{minfit}
\end{figure}

In order to estimate the most probable deuteron beam energy and associated uncertainty, a parabola is fit to the best-fit chi-square statistic as a function of simulated beam energy, as shown in Figure~\ref{minfit}. The value of the energy at the minimum, $S_\mathrm{min}$, is used as the best estimate of the beam energy parameter and the corresponding uncertainty is extracted from the width of the parabola at a value of $S_\mathrm{min} +1$ \cite{Lyons}.

The best fit deuteron beam energy for the entire dataset constrains deuteron beams to $7.4 \pm 5.6^{(\mathrm{stat})} \pm 3.7^{(\mathrm{syst})}$~keV, where stat. and syst. denote the statistical and systematic contributions to the error, described in section \ref{Systematics} below.

\subsection{Time-resolved 25~kV result}

\begin{figure}
     \centering
     \begin{subfigure}[b]{0.45\textwidth}
         \centering
         \includegraphics[width=\textwidth]{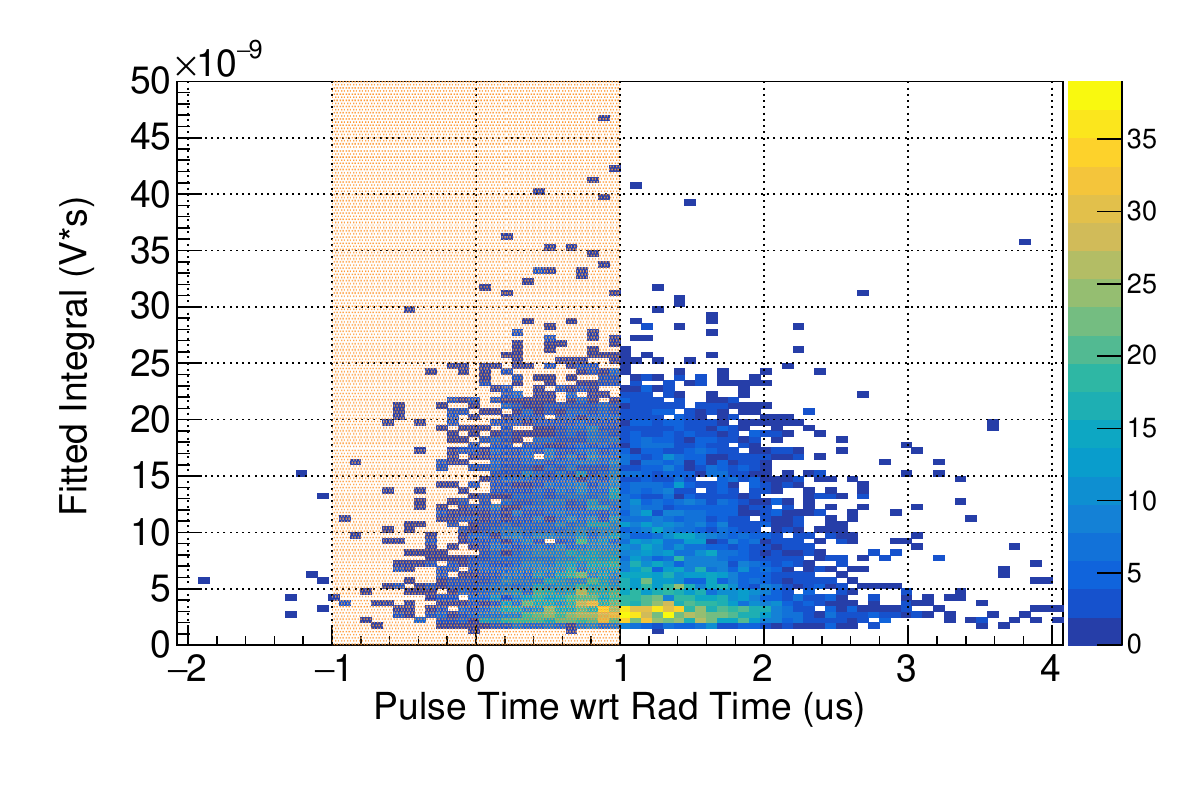}
         \caption{}
         \label{tdep_windows}
         
     \end{subfigure}
     \hfill
     \begin{subfigure}[b]{0.5\textwidth}
         \centering
         \includegraphics[width=\textwidth]{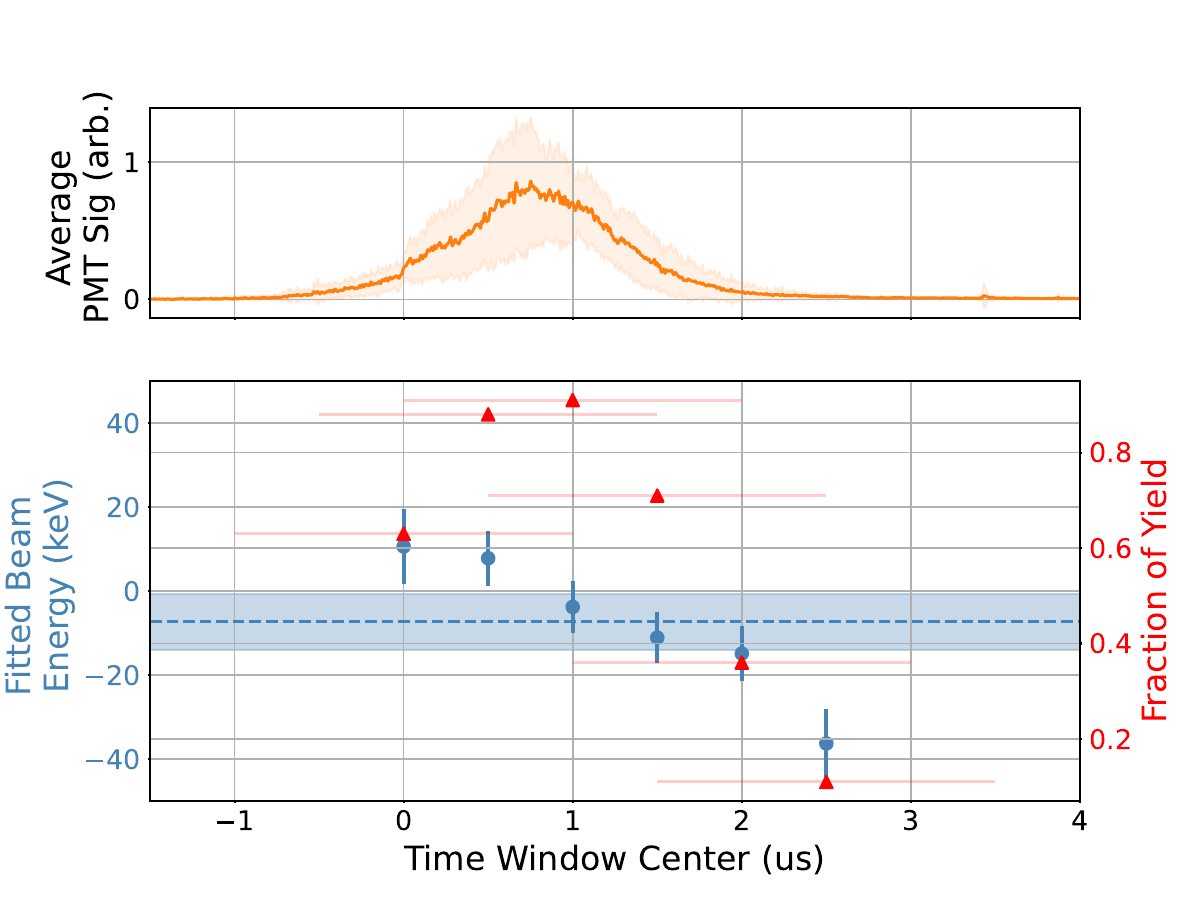}
         \caption{}
         \label{enVsYieldVsTime}
     \end{subfigure}
     \caption{(a) Two dimensional histogram showing the number of pulses observed in time relative to radiation start versus the pulse energy. Shown in orange is a sliding 2~{\textmu s} window in which all pulse energies are included in the forward-fit analysis for each time slice, which is advanced in $0.5$~{\textmu s} intervals. (b) The results of the best fit procedure to infer beam energy for the sliding 2~{\textmu s} time window. The fitted deuteron beam energy $E_d$ is shown in blue and the fraction of the total neutron yield that occurs in each 2~{\textmu s} window is displayed in red. The red bands on each point for the yield fraction indicate the 2~{\textmu s} fit window and the mean PMT signal as a function of time is shown in orange. The blue dashed line and shaded band denotes the fitted result and associated statistical error for the entire dataset. For the majority of the radiation event, the measured beam energy is within the uncertainty band of the average taken over the full dataset. The beam energy in the final window, centered at 2.5~us shows an increase of more than a factor of two in magnitude but contains less than 10\% of the neutron signal contributing to the total yield.}
     \label{time_res}
\end{figure}

For each shot, signals from an axial array of scintillator detectors operating in current mode are used to define a radiation start time as the time at which 20\% of the maximum signal is reached. Alignment of all discharges with respect to the radiation start time allows for a time resolved isotropy result. Figure~\ref{tdep_windows} shows the number of pulses detected by the isotropy detectors across the full range of pulse energies and observation times relative to radiation start. 
The shaded region denotes a 2~{\textmu s} time window which is shifted in $0.5$~{\textmu s} increments to produce pulse integral spectra for a series of time windows within the radiation event. Figure~\ref{enVsYieldVsTime} displays the resulting beam energy obtained from the forward-fit method for each time window as well as the fraction of the total neutron yield contained within each. For windows later in time, the best fit is consistent with a slightly increased deuteron beam energy. The most significant effect is seen in the latest time window, starting at $1.5$~{\textmu s}. Here an increase in the magnitude of the beam energy is observed, hinting at an instability near the end of the radiation event. However, only 10\% of the total yield is produced within this time window, suggesting the majority of the neutrons produced over the entire shot are not produced via beam-target interactions.

\subsection{Evaluation of systematic errors} \label{Systematics}

Pulse pileup, where nearly-simultaneous recoil events produce indistinguishable pulses, as well as gamma radiation can both contaminate the energy spectra with signals not originating from individual neutron interactions within the scintillator.  The significance of these sources of systematic error to the isotropy analysis are quantified in detail.

\subsubsection{Pulse pileup}

The increased neutron yields produced by recent operating conditions increase the probability of pulse pileup. Unresolved pileup events inflate the integrated energy of a given scintillation pulse. If enough pulse energies are misinterpreted due to this effect, the edge location of the pulse integral could shift to higher energies skewing the fitted value of the isotropy. In order to quantify this effect, two key values must be determined: the fraction of measured pulses which contain an unresolved second pulse, and the shift of the interpreted deuteron beam energy given a specific fraction of piled-up pulses.

\begin{figure}
    \centering
    \includegraphics[width=0.7\linewidth]{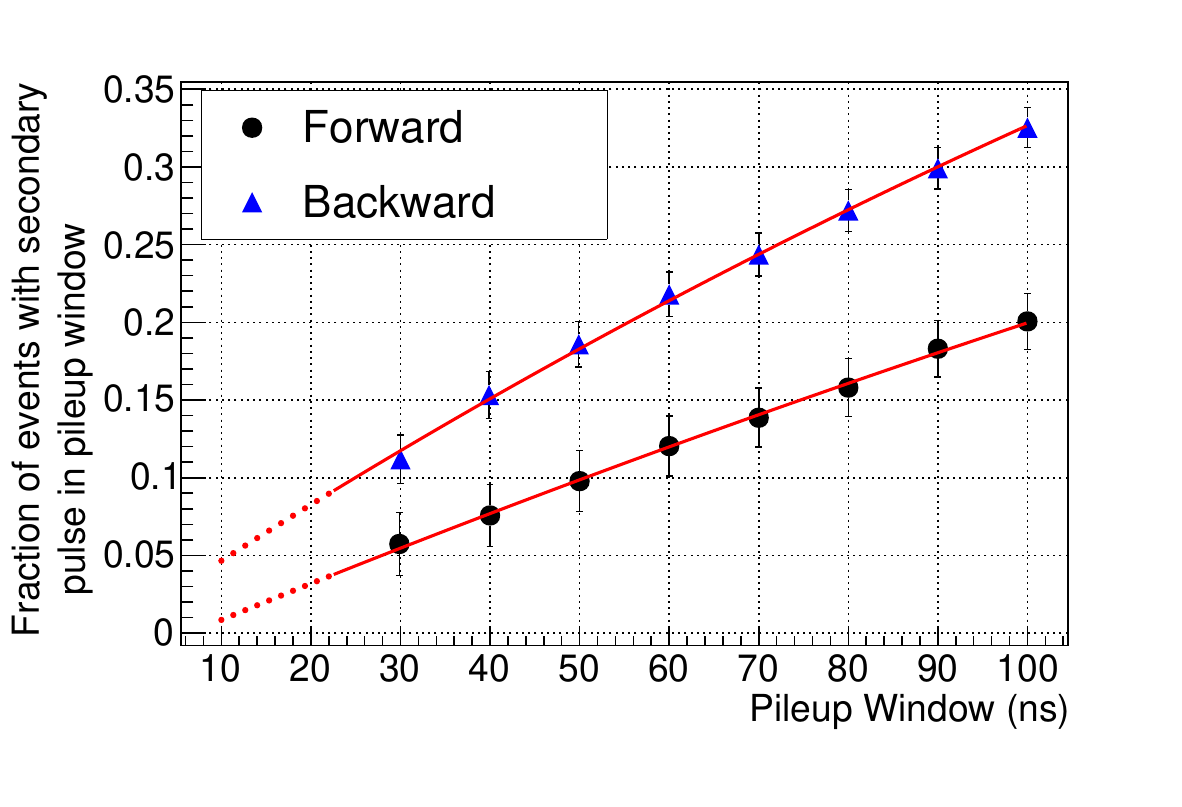}
    \caption{Fraction of pulses with a secondary event in a given time window, $\tau$, as a function of the time window. The fit to each detector is used to extrapolate back to 10~ns, where two adjacent pulses are unresolvable in the analysis, to estimate the amount of pileup in the dataset. }
    \label{puextrap}
\end{figure}

Pulses separated by less than 10~ns are not resolved as individual pulses within the analysis, a constraint imposed by time response of the scintillator and photomultiplier. The number of secondary pulses which arrive after this 10~ns period can be quantified and an extrapolation can be used to estimate the probability of an unresolvable pileup event. The fraction of events in which a secondary pulse arrives within a given time window of a preceding pulse as a function of the time window can be seen in Figure~\ref{puextrap}. The backward detector is located 25~cm closer to the center of the device, leading to an increased probability of pileup. A fit and extrapolation to 10~ns estimates the fraction of unresolved pulses in the forward and backward detector sets to be 0.85\% and 4.6\%, respectively. 

\begin{figure}
     \centering
     \begin{subfigure}[b]{0.45\textwidth}
         \centering
         \includegraphics[width=\textwidth]{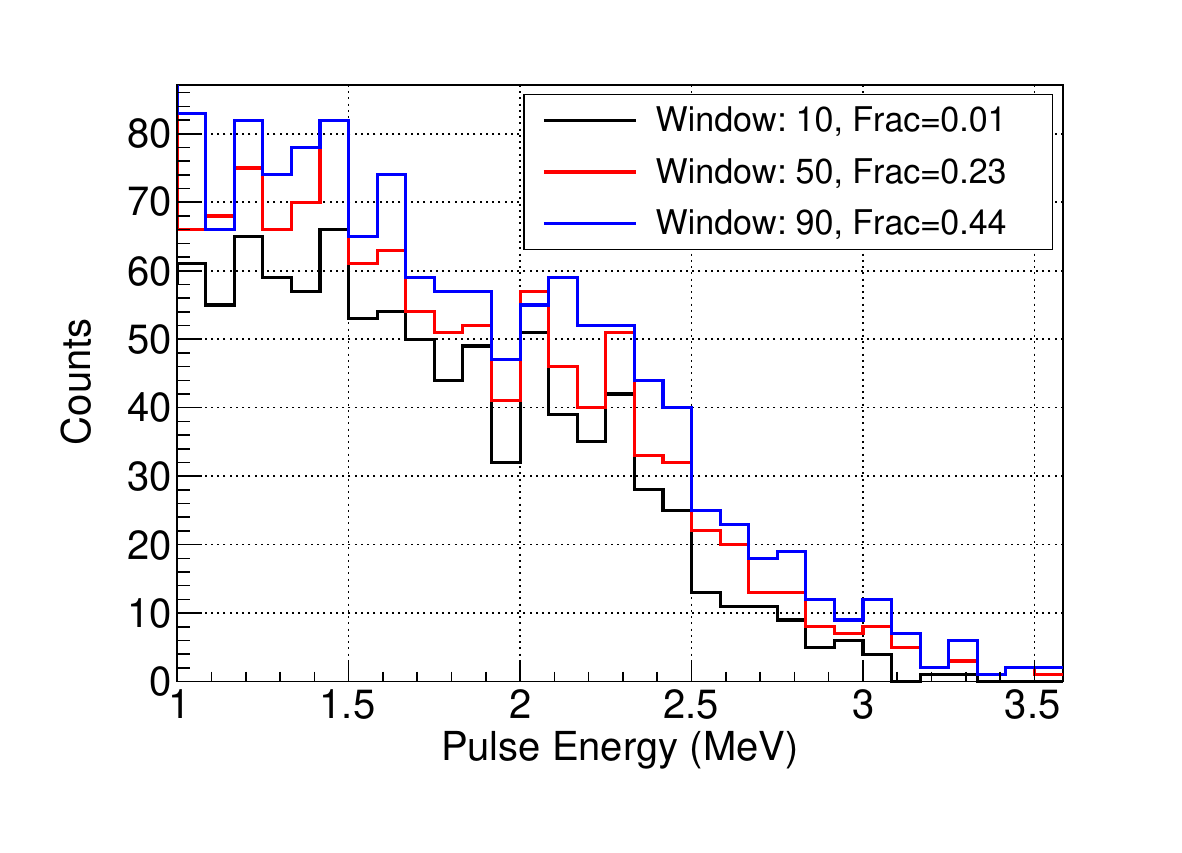}
         \caption{}
         \label{pis_pu}
     \end{subfigure}
     \hfill
     \begin{subfigure}[b]{0.45\textwidth}
         \centering
         \includegraphics[width=\textwidth]{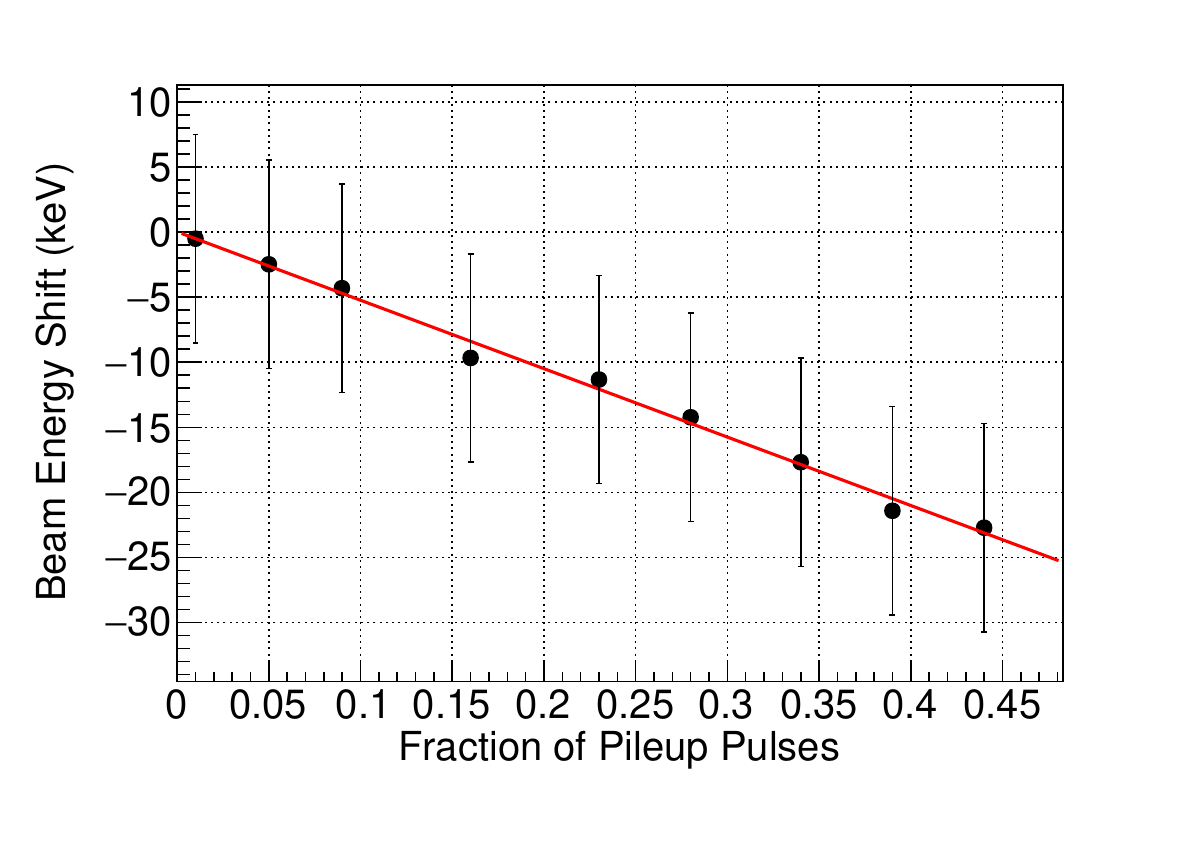}
         \caption{}
         \label{pileupShift}
     \end{subfigure}
     \caption{Method used to quantify the effect of pileup pulses on the final isotropy result. (a) Pulse energy distributions for the full dataset varying the fraction of artificially summed pulses to mimic pulse pileup show the shift of the pulse integral edge location to larger values of energy. (b) Shift in the fitted beam energy as a function of the fraction of pulses which are piled up obtained by running the shifted pulse integral spectra through the simultaneous fit routine. The fraction of unresolved pulses is estimated to be 0.046, implying a systematic uncertainty of 2.5~keV for the beam energy measurement based on this curve.}
     \label{pileup_plots}
\end{figure}

To characterize the shift of the fitted beam energy from a given fraction of unresolved, overlapping pulses, pulse integral spectra were produced from experimental data by summing adjacent pulses within specified time windows. The resulting spectra contain known quantities of summed pulse energies. As shown in Figure~\ref{pileup_plots}, as the window is increased from 30 to 100~ns, the fraction of pulses which have been summed together increases, simulating increased pileup, and the edge of the pulse integral spectrum shifts to higher energy. 
Repeating the forward-fit procedure with the artificially shifted pulse integral spectrum for one detector location and the original spectrum at the second location results in the inference of maximal beam energy attributable to pileup for a given pileup fraction, shown in Figure~\ref{pileup_plots}b.
Given that the limiting fraction of unresolved pulses at the backwards detector location is estimated to be 4.6\%, the systematic uncertainty associated with pulse pileup is estimated as $2.5$~keV.

\subsubsection{Gamma background} \label{sec:gamma_cont}

\begin{figure}
\centering
 \includegraphics[width=\linewidth]{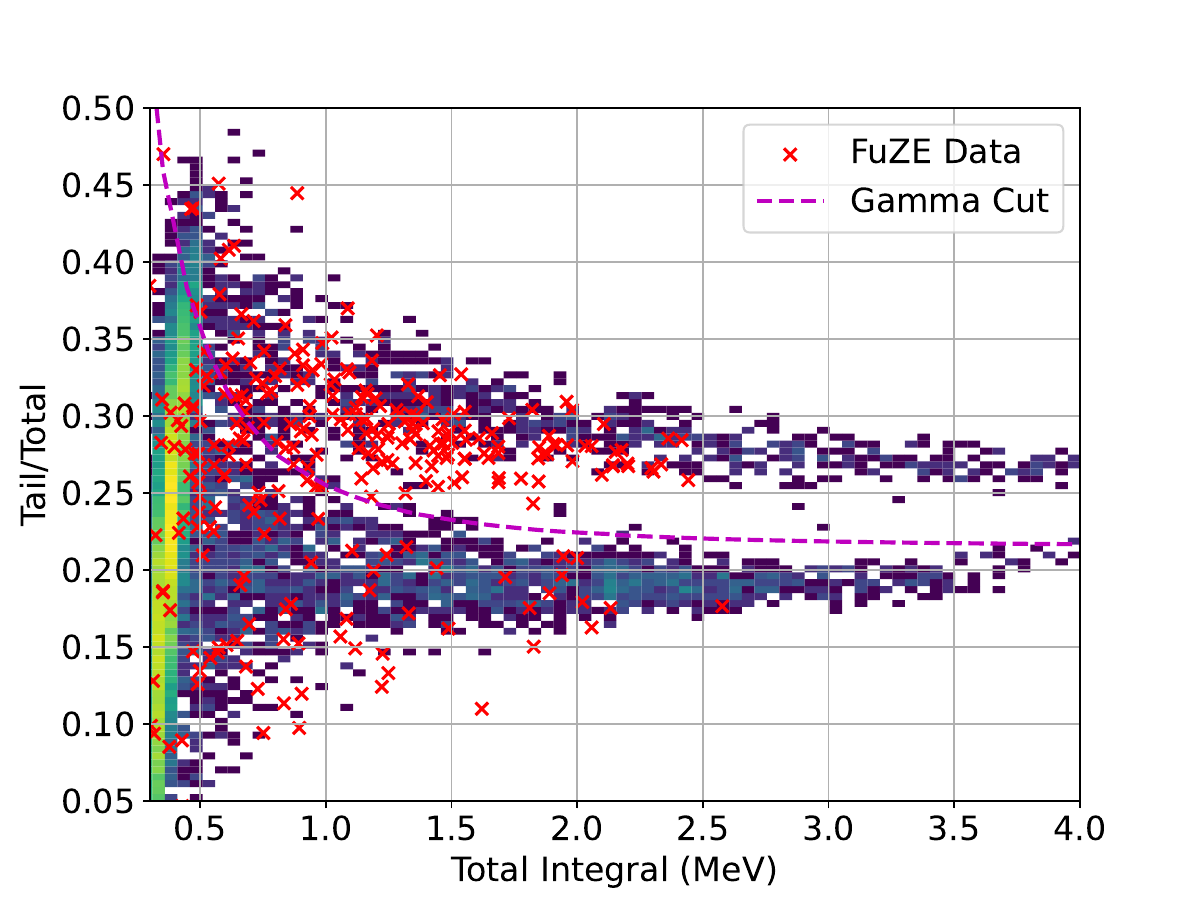}
 \caption{The red $\times$ symbols denote the PSD parameters for experimental pulses taken during the FuZE campaign for a single organic glass detector. The magenta dashed line shows the three sigma band used to distinguish neutron pulses from gamma pulses. The experimental data is plotted on top of the AmBe calibration as a visual aid. Pulses within the range of the forward-fit from 1.5 to 3 MeV are in a region of good neutron/gamma separation capability.}
 \label{psd}
\end{figure}

The signal decay of organic scintillators consists of prompt and delayed components. The technique of using the relative intensities of the two components to discriminate between particle types has been well established and documented \cite{Brooks}, and is only briefly described here. Neutron interactions in the scitnillator produce an enhanced time delayed component and can be distinguished from gamma interactions by quantifying the fraction of total charge contained in the tail of the pulse. This tail-to-total ratio is the pulse shape discrimination (PSD) parameter, and is shown versus the total pulse integral for an americium-beryllium (AmBe) source in the color plot of Figure~\ref{psd}. Neutrons and gammas populate the upper and lower bands, respectively. The dashed line shows the energy dependent three sigma discrimination of gamma particles. 

The PSD parameter versus total pulse energy observed for all pulses in a single organic glass scintillator across 400 FuZE discharges are displayed as red $\times$ symbols in Figure~\ref{psd}. The data is overlayed with the AmBe calibration plot to visualize the distribution of neutron and gamma pulses, and the dashed line displays the energy dependent three sigma discrimination between particle types. Within the range of the energies included in the forward-fit, all pulses are above the threshold where PSD separation of the two particle types is possible. 

\begin{figure}
    \centering
    \includegraphics[width=\linewidth]{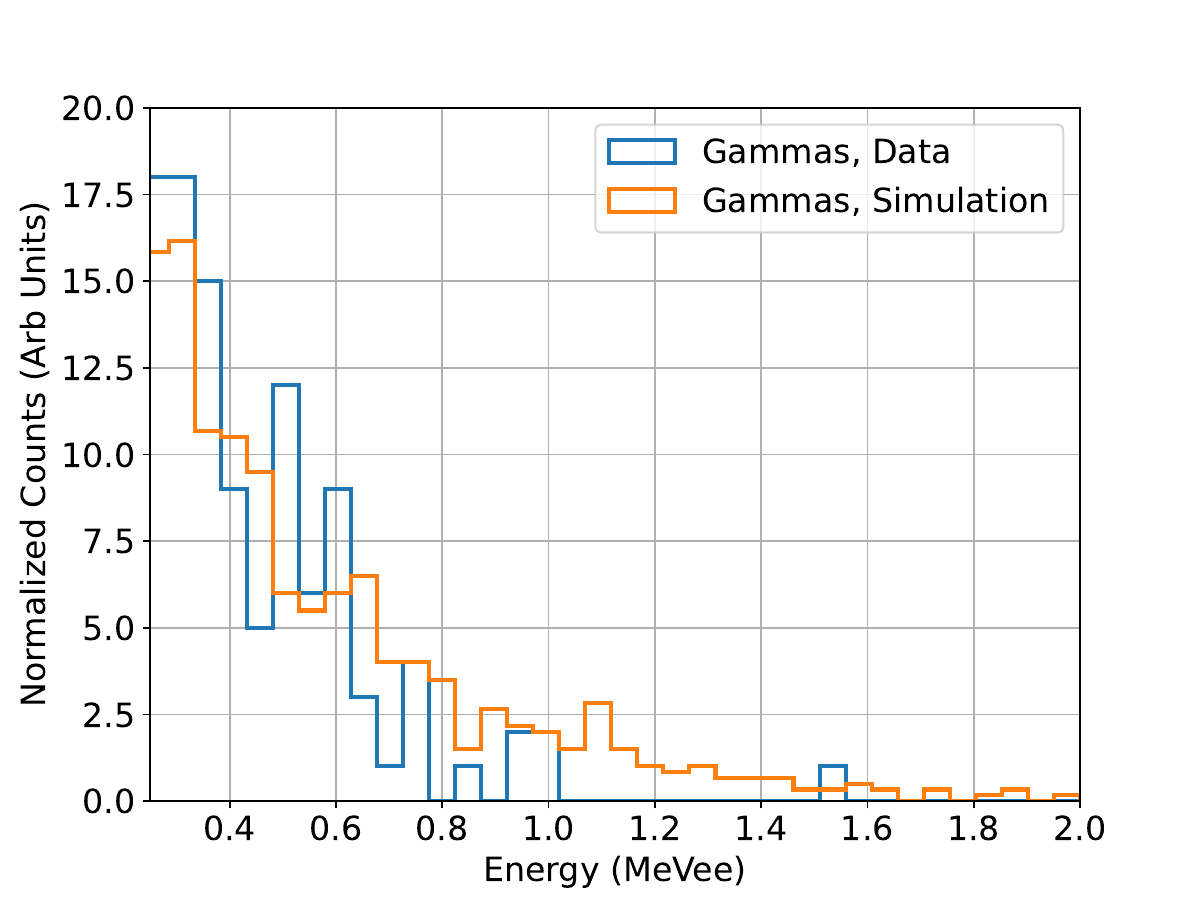}
    \caption{Comparison between the simulated gamma energy distribution and the measured gamma energy distribution. The simulated data is scaled to compare to the data. Although statistically limited by the total number of gamma counts observed, the measured distribution does not deviate significantly from the simulated distribution.}
    \label{gamma_Comp}
\end{figure}

The objective of this analysis is to determine whether the gamma background simulated by MCNP adequately describes the observed gamma background. Experimentally observed PSD pulses included in the forward-fit were found to have a gamma-to-neutron ratio of $0.25 \pm 0.04$, whereas the simulated gamma-to-neutron ratio for comparable pulse energies is $0.13 \pm 0.01$. Although statistically limited by the number of counts observed in total, the shape of the gamma energy distribution does not appear to deviate significantly between simulation and data, as shown in Figure~\ref{gamma_Comp}.

In order to quantify the effect of a two-fold increase in the gamma contribution on the final isotropy measurement, the fit procedure outlined in Section \ref{fit_analysis} was repeated while varying the magnitude of the gamma contribution. 
Gamma backgrounds 2--5 times the original simulated gamma background result in fits with greater chi-squared statistic and a corresponding increase in uncertainty, but there is little effect on the beam energy inferred by the minimization.

The maximum shift in inferred beam energy attributable to the increased magnitude gamma background was roughly 2~keV, which we assign as the overall systematic uncertainty due to the gamma background.

\section{Conclusion} \label{Conclusion}
Measurements of the neutron energy at two locations have been used to quantify deuteron beam energy for increasingly greater neutron yields observed on the FuZE device during a 433 shot campaign. By fitting the data recorded at two different locations to simulated spectra, the deuteron beam energy is determined to be less than $7.4 \pm 5.6^{(\mathrm{stat})} \pm 3.7^{(\mathrm{syst})}$~keV. Pulse shape discriminating detectors have been implemented to validate the simulated gamma contributions, and the systematic error associated with pulse pileup has been determined.

Time-resolved measurements reveal an increased beam-target contribution at late times, suggesting the presence of potential instability activity towards the end of the radiation event. However, the majority of the neutron yield is produced during a period of neutron energy isotropy.

An upgraded version of the detector system presently in development will deploy more scintillators per array, reducing the required number of discharges to establish comparably low statistical uncertainty. This upgraded detector array and the analysis presented here will be implemented on FuZE-Q, a SFS Z-pinch device capable of producing significantly higher neutron yields per discharge.
\\ \\

\bibliographystyle{unsrt} 
\bibliography{refs} 

\begin{thebibliography}{10}

\bibitem{Anderson}
Oscar~A. Anderson, William~R. Baker, Stirling~A. Colgate, John Ise, and Robert~V. Pyle.
\newblock {Neutron Production in Linear Deuterium Pinches}.
\newblock {\em Phys. Rev.}, 110:1375--1387, Jun 1958.

\bibitem{Kadomtsev}
BB~Kadomtsev et~al.
\newblock Hydromagnetic stability of a plasma.
\newblock {\em Reviews of plasma physics}, 2:153--199, 1966.

\bibitem{Shumlak_Hartman}
U.~Shumlak and C.~W. Hartman.
\newblock {Sheared Flow Stabilization of the $\mathit{m}\phantom{\rule{0ex}{0ex}}=\phantom{\rule{0ex}{0ex}}1$ Kink Mode in $\mathit{Z}$ Pinches}.
\newblock {\em Phys. Rev. Lett.}, 75:3285--3288, Oct 1995.

\bibitem{Shumlak:2001}
U.~Shumlak, R.~P. Golingo, B.~A. Nelson, and D.~J. Den~Hartog.
\newblock {Evidence of Stabilization in the $\mathit{Z}$-Pinch}.
\newblock {\em Phys. Rev. Lett.}, 87:205005, Oct 2001.

\bibitem{ShumlakJAP2020}
U.~Shumlak.
\newblock {{Z-pinch fusion}}.
\newblock {\em Journal of Applied Physics}, 127(20):200901, 05 2020.

\bibitem{Zhang2019}
Y.~Zhang, U.~Shumlak, B.~A. Nelson, R.~P. Golingo, T.~R. Weber, A.~D. Stepanov, E.~L. Claveau, E.~G. Forbes, Z.~T. Draper, J.~M. Mitrani, H.~S. McLean, K.~K. Tummel, D.~P. Higginson, and C.~M. Cooper.
\newblock {Sustained Neutron Production from a Sheared-Flow Stabilized $Z$ Pinch}.
\newblock {\em Phys. Rev. Lett.}, 122:135001, Apr 2019.

\bibitem{Levitt2024PRL}
B.~Levitt, C.~Goyon, J.~T. Banasek, S.~C. Bott-Suzuki, C.~Liekhus-Schmaltz, E.~T. Meier, L.~A. Morton, A.~Taylor, W.~C. Young, B.~A. Nelson, D.~A. Sutherland, M.~Quinley, A.~D. Stepanov, J.~R. Barhydt, P.~Tsai, K.~D. Morgan, N.~van Rossum, A.~C. Hossack, T.~R. Weber, W.~A. McGehee, P.~Nguyen, A.~Shah, S.~Kiddy, M.~Van~Patten, A.~E. Youmans, D.~P. Higginson, H.~S. McLean, G.~A. Wurden, and U.~Shumlak.
\newblock {Elevated Electron Temperature Coincident with Observed Fusion Reactions in a Sheared-Flow-Stabilized $Z$ Pinch}.
\newblock {\em Phys. Rev. Lett.}, 132:155101, Apr 2024.

\bibitem{Mitrani:2021}
James~M. Mitrani, Joshua~A. Brown, Bethany~L. Goldblum, Thibault~A. Laplace, Elliot~L. Claveau, Zack~T. Draper, Eleanor~G. Forbes, Ray~P. Golingo, Harry~S. McLean, Brian~A. Nelson, Uri Shumlak, Anton Stepanov, Tobin~R. Weber, Yue Zhang, and Drew~P. Higginson.
\newblock {{Thermonuclear neutron emission from a sheared-flow stabilized Z-pinch}}.
\newblock {\em Physics of Plasmas}, 28(11):112509, 11 2021.

\bibitem{TechReport_MCNP}
Christopher~John Werner, Jerawan~Chudoung Armstrong, Forrest~Brooks Brown, Jeffrey~S. Bull, Laura Casswell, Lawrence~James Cox, David~A. Dixon, Robert~Arthur Forster, III, John~Timothy Goorley, Henry~Grady Hughes, III, Jeffrey~A. Favorite, Roger~Lee Martz, Stepan~Georgievich Mashnik, Michael~Evan Rising, Clell~Jeffrey Solomon, Jr., Avneet Sood, Jeremy~Ed Sweezy, Anthony~J. Zukaitis, Casey~Alan Anderson, Jay~Samuel Elson, Joe~W. Durkee, Jr., Russell~Craig Johns, Gregg~Walter McKinney, Garrett~Earl McMath, John~S. Hendricks, Denise~B. Pelowitz, Richard~Edward Prael, Thomas~Edward Booth, Michael~R. James, Michael~Lorne Fensin, Trevor~A. Wilcox, and Brian~Christopher Kiedrowski.
\newblock {MCNP User's Manual Code Version 6.2}.
\newblock Technical Report LA-UR-17-29981, Los Alamos National Laboratory, Los Alamos, NM, USA, October 2017.

\bibitem{CARLSON2016}
Joseph~S. Carlson and Patrick~L. Feng.
\newblock {Melt-cast organic glasses as high-efficiency fast neutron scintillators}.
\newblock {\em Nuclear Instruments and Methods in Physics Research Section A: Accelerators, Spectrometers, Detectors and Associated Equipment}, 832:152--157, 2016.

\bibitem{SICILIANO2008}
E.R. Siciliano, J.H. Ely, R.T. Kouzes, J.E. Schweppe, D.M. Strachan, and S.T. Yokuda.
\newblock {Energy calibration of gamma spectra in plastic scintillators using Compton kinematics}.
\newblock {\em Nuclear Instruments and Methods in Physics Research Section A: Accelerators, Spectrometers, Detectors and Associated Equipment}, 594(2):232--243, 2008.

\bibitem{DIETZE1981}
G.~Dietze and H.~Klein.
\newblock {Gamma-calibration of NE 213 scintillation counters}.
\newblock {\em Nuclear Instruments and Methods in Physics Research}, 193(3):549--556, 1982.

\bibitem{pozzi}
Sara~A. Pozzi, James~A. Mullens, and John~T. Mihalczo.
\newblock {Analysis of neutron and photon detection position for the calibration of plastic (BC-420) and liquid (BC-501) scintillators}.
\newblock {\em Nuclear Instruments and Methods in Physics Research Section A: Accelerators, Spectrometers, Detectors and Associated Equipment}, 524(1):92--101, 2004.

\bibitem{weldon}
R.A. Weldon, J.M. Mueller, P.~Barbeau, and J.~Mattingly.
\newblock {Measurement of EJ-228 plastic scintillator proton light output using a coincident neutron scatter technique}.
\newblock {\em Nuclear Instruments and Methods in Physics Research Section A: Accelerators, Spectrometers, Detectors and Associated Equipment}, 953:163192, 2020.

\bibitem{manfredi}
J.~J. Manfredi, B.~L. Goldblum, T.~A. Laplace, G.~Gabella, J.~Gordon, A.~O’Brien, S.~Chowdhury, J.~A. Brown, and E.~Brubaker.
\newblock {Proton Light Yield of Fast Plastic Scintillators for Neutron Imaging}.
\newblock {\em IEEE Transactions on Nuclear Science}, 67(2):434--442, 2020.

\bibitem{Lyons}
L.~Lyons.
\newblock {\em {{S}tatistics for nulcear and particle physicists}}.
\newblock Cambridge University Press, 1986.

\bibitem{Brooks}
F.D. Brooks.
\newblock {Development of organic scintillators}.
\newblock {\em Nuclear Instruments and Methods}, 162(1):477--505, 1979.

\end{thebibliography}

\end{document}